# Factors that Contribute to the Success of a Software Organisation's DevOps Environment: A Systematic Review

Ashley Gwangwadza[1] and Ridewaan Hanslo[2][0000-0003-4075-722X]

[1] University of Pretoria, Gauteng, South Africa
`ashley.gwangwadza@up.ac.za`
[2] University of Pretoria, Gauteng, South Africa
`ridewaan.hanslo@up.ac.za`

**Abstract.** This research assesses the aspects of software organisations' DevOps environments and identifies the factors contributing to these environments' success. DevOps is a recent concept, and many organisations are moving from old-style software development methods to agile approaches such as DevOps. However, there is no comprehensive information on what factors impact the success of the DevOps environment once organisations adopt it. This research focused on addressing this gap through a systematic literature review. The systematic review consisted of 33 articles from five selected search systems and databases from 2015 to 2021. Based on the included articles, 15 factors were identified and grouped into four categories: Collaborative Culture, Organizational Aspects, Tooling and Technology, and Continuous Practices. In addition, this research proposes a DevOps environment success factors model to potentially contribute to DevOps research and practice. Recommendations are made for additional research on the effectiveness of the proposed model and its success factors.

## 1 Introduction

Digital transformation presents new challenges for software organisations with the increasing demand for customised IT solutions [23]. Furthermore, to sustain a competitive advantage, software organisations need to deliver valuable and quality IT solutions in short cycles [30]. To address this need, many software organisations implement continuous practices: continuous delivery, continuous deployment and continuous integration [1]. In addition to continuous practices, recent studies show that DevOps adoption also improves delivery cycle times and quality [30].

The term DevOps comes from the terms of two of the core software development teams: development and operations [6]. There is no clear definition for the DevOps term. Still, the most shared interpretation is DevOps is a set of practices based on a culture that emphasises integration and collaboration between the operations teams, the developer teams, and all persons who participate in the design, development, and deployment of software [27].

DevOps takes a continuous approach that extends outside the borders of old-style software development and extends to operations [5]. Thus, DevOps implementation has been increasing as more software organisations are opting to move from the old-style software development approaches such as the Waterfall Model and Spiral Model to agile approaches as a way to acquire a competitive advantage [32]. The competitive advantage stems from the benefits of DevOps, which include a shorter time-to-market cycle, improved customer satisfaction, greater developer efficiency and productivity, continuous quick feedback and higher software quality and reliability [5].

However, adopting DevOps remains challenging because although there is an excess of information, DevOps tools and DevOps practices, there is a lack of strategy to organise the information for correct DevOps adoption [21]. The adoption of DevOps affects different aspects of a software organisation such as organisational structures, company culture, products, processes and related technologies used in software development [19]. Thus, many factors such as constraints from the environments and the lack of clarity of the meaning of the term DevOps may lead to DevOps environments not always being successful [22].

Therefore, the research problem is that there is comprehensive information available on tools, practices, benefits, and challenges of DevOps but no comprehensive information for software organisations on the factors that contribute to a successful DevOps environment. Because of this research gap, it is poised that there is a need to research the factors contributing to the success of DevOps environments within a software organisation context.

This research aims to identify and describe the factors that contribute to the success of DevOps environments within software organisations. This is done through a systematic literature review.

The paper is structured as follows; section 2 explains the research method. Section 3 details the results of this study and section 4 discusses the findings. Section 5 concludes the research, providing research implications, limitations and recommendations for future research.

## 2  Research Method

The research is conducted using a systematic literature review (SLR), defined as "*a means of identifying, evaluating and interpreting all available research relevant to a particular research question, or topic area, or phenomenon of interest*" [15]. Simply put, an SLR is a review of primary studies. This study follows the SLR guidelines by [15], which are: identifying resources, study selection, data extraction, data synthesis and writing up the study as a report.

### 2.1 Research Question

This research aims to answer the following question:
What are the factors that contribute to the success of DevOps environments within software organisations?

### 2.2 Search Terms

"DevOps" AND ("continuous deployment" OR "continuous development" OR "environments" OR "development" OR "software engineering" OR "methodology" OR "software organisation" OR "software" OR "factors" OR "success")

### 2.3 Selection Criteria

The selection of research material for inclusion in this systematic review is based on the following inclusion and exclusion criteria.

**Inclusion Criteria.** For a source to be included in the research, it had to meet the following criteria:
- Papers that identify the factors that influence a DevOps Environment
- Papers that discuss DevOps environments within a software organisation
- Papers that discussed practices or factors that contribute to the success of DevOps environments
- Journal articles, conference papers, book chapters, dissertations and thesis are considered for review
- Papers containing at least three keywords in the title, abstract, keywords

**Exclusion Criteria.** Sources are excluded from the research for the following reasons:
- Papers with no explicit discussion about DevOps
- Papers that do not provide information on factors contributing to DevOps success
- The full text of a paper is not available
- The paper is not written in English
- Duplicate papers (the same paper taken from different databases)
- Duplicate reports of the same study (only the most complete version is included)

### 2.4 Source Selection

The following data sources were selected to perform the search:
a) IEEE Xplore Digital Library
b) Science Direct
c) ACM Digital Library
d) Springer Link

These databases are all recognised research databases in the information technology field. In addition, Google Scholar was used to find additional sources missing in the abovementioned databases. This study uses the words article and paper interchangeably to reference the data source. However, the data source is either a journal article, conference paper, book chapter, dissertation, or thesis.

Figure 1 displays a Prisma Flowchart with four steps, Identification, Screening, Eligibility, and Included. Sources are excluded during each step as depicted in the flowchart based on the predefined selection criteria detailed in section 2.3.

A search using the search string mentioned in section 2.2 was performed on the selected databases, returning 2179 articles. The Google Scholar citation search found an additional 10 records. After that, 291 duplicate papers were removed. Screening by the title was conducted, leaving 262 full-text articles. An abstract screening was then performed on these articles, leaving 108. After the full-text assessment for eligibility, the 33 remaining articles were used for data extraction and synthesis.

### 2.5    Prisma Flowchart

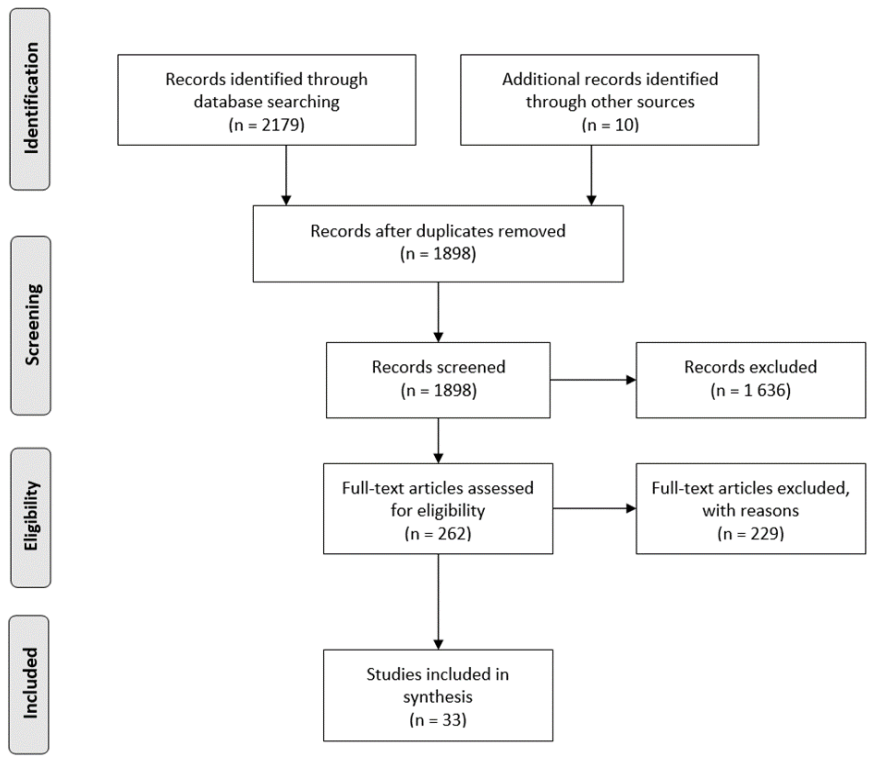

**Fig. 1.** Prisma Flowchart

### 2.6 Quality Assessment

The included papers were assessed using four quality assessment questions. The questions aimed to evaluate the quality aspects mentioned by [15]. These aspects are characterised as **objectivity** - if the research is free of bias; **reliability** - the accuracy and reliability of the research instruments used; **internal validity** - whether the research was well structured, so data was collected from suitable sources, and **external validity** - determines if the findings can be predicted for subsequent occasions.

Therefore, the following questions were devised to assess the quality of the selected literature:
Q1. Are DevOps and the DevOps environments the core of the discussion?
Q2. Does the research have a well-defined aim?
Q3. Does the article follow a clear research process with a clear description of the methods used to analyse data?
Q4. Does the article report its findings based on evidence and argument?

The response options to these questions were Yes, No and Partially. The weighting for each response is as follows: Yes = 1, Partially = 0.5, No = 0. The total score was then recorded and used as a guide to indicate the quality of the selected literature ranging from 0 to 4. The quality evaluation of the articles is described in section 3.2

### 2.7 Data Extraction

The data extraction was carried out on the 33 papers included in the systematic literature review. After that, a qualitative thematic analysis was conducted to synthesise the extracted data. While reading each paper, sections of its text were highlighted to identify the concepts. These concepts, also known as codes, were closely examined to categorise them into common themes. All the relevant information that helped answer the research question was extracted, including the citation, the research title, the source database, the year published and study type, article sub-concepts, and the central concept. Microsoft Excel was used for data extraction and synthesis.

## 3 Results

An analysis of the extracted data was conducted, which includes information on the search results, quality evaluation of the included papers, and the synthesis of identified factors.

### 3.1 Search Results

According to [23], the first publications devoted to DevOps originated in 2011. However, in this study, the earliest publications on the factors that influence a DevOps environment were published in 2015. Figure 2 depicts the distribution of the included papers that satisfied the selection criteria of this study.

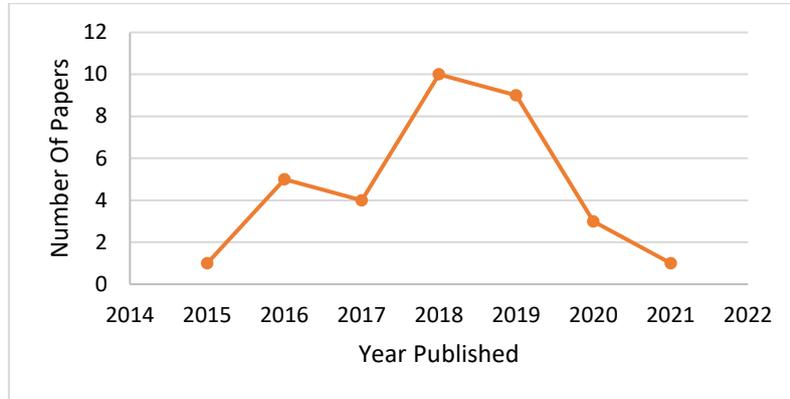

**Fig. 2.** Distribution of selected literature by publication year

### 3.2 Quality Evaluation of Articles

Four questions were used to assess the quality of the selected literature, as mentioned in the quality assessment section. The majority of the included papers had a quality assessment score of 4 (76%), with the remaining papers having a score between 2.5 and 3.5 (24%).

### 3.3 Synthesis of Identified Factors

A thematic analysis was conducted to identify the core themes and factors within the selected literature. The themes corresponded with the constructs that held the factors identified. Initially, 16 factors were identified as contributing to the success of the DevOps environment. However, three factors merged into others, leaving 13 factors. The 13 factors result from re-examining each of these factors' definitions and the references made by the authors of the selected literature. The factors that were merged into others were Cooperation, Coordination and Knowledge sharing. Although none of the authors clearly defined Cooperation and Coordination, they were referenced in relation to collaboration by [29] and sometimes used in place of the word collaboration [22]. Therefore, Cooperation and Coordination were merged into the Collaboration factor. Knowledge management was found in only two papers and was defined by [5] as collection, visualisation and sharing of information and knowledge to support collaboration. Thus, this factor was then merged into the Sharing factor, which includes knowledge sharing and had more existing literature supporting it.

The remaining 13 factors were then analysed to identify any relationships that could help categorise them. Categorising the factors would help to understand the more significant themes and give deeper insight into the factors. Four common themes were identified: Collaborative Culture, Organisational Aspects, Tooling and Technology, and Continuous Practices, as illustrated in Table 1.

In addition, a qualitative data analysis tool called NVivo was used to refine the grouping of the factors and assess the validity of the selected themes. All the data extracted was processed in NVivo, and it generated 19 codes. Several of these codes corresponded with the initially identified thematic analysis factors. This tool grouped the codes into categories, and the result was that NVivo also grouped the factors as they were grouped in the thematic analysis. Albeit, NVivo gave a few more codes than initially identified in the first iteration of the data syntheses. After analysing and considering each of these codes as possible factors, two additional codes were included as part of the synthesised data, resulting in a total of 15 factors (see Table 1). These two additional codes are Continuous Deployment and Continuous Delivery. The rest of the codes were excluded as they were ambiguous or it did not contribute to answering the research question. The DevOps success factors conceptual model is shown in Figure 3, depicting the core themes and factors.

**Table 1.** Identified DevOps Environment Success Factors summary

| Success Factor | Sources |
| --- | --- |
| 1. Communication | [2], [16], [24, 25], [31] |
| 2. Collaboration | [2], [6], [9, 10], [12], [14], [17], [20], [23], [28, 29], [32, 33] |
| 3. Sharing | [1], [4-9], [11], [17], [22], [28] |
| 4. Team Roles Setting | [6], [10], [19], [21], [26] |
| 5. Organisational Culture | [1], [4, 5], [7], [9], [12, 13], [17], [19], [25], [31] |
| 6. Training | [2], [16], [31] |
| 7. Customer Involvement | [1], [5], [7], [17], [24] |
| 8. Automation | [1], [3-9], [11], [17], [19], [22], [28], [32] |
| 9. Measurement | [1], [3-5], [9], [17], [19], [23], [28] |
| 10. Tools | [7], [13], [16], [19], [22] |
| 11. Continuous Integration | [8], [14], [19], [27], [29, 30] |
| 12. Continuous Experimentation | [3], [11], [14], [17], [19], [22], [27], [30] |
| 13. Continuous Delivery | [14], [17], [30] |
| 14. Continuous Deployment | [3], [11], [14], [19], [27], [30], [32] |
| 15. Continuous Monitoring | [8], [11], [19], [22] |

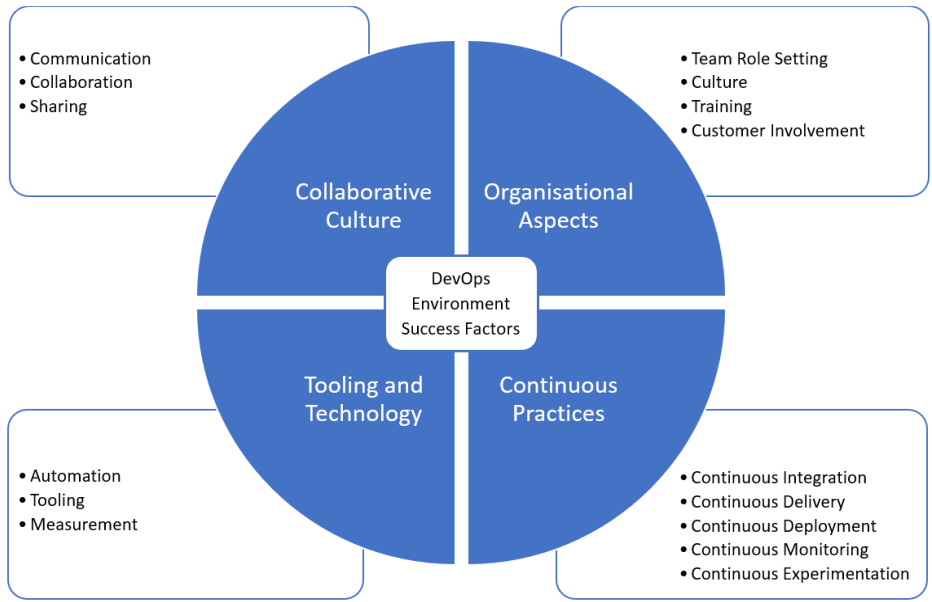

**Fig. 3.** DevOps Environment Success Factors Model

## 4 Discussion

This section aims to answer the research question. The core themes identified are 1) Collaborative Culture; 2) Organisational Aspects; 3) Tooling and Technology, and 4) Continuous Practices. Each of these themes has three to five factors. Therefore, the discussion will first explain how these themes were formed and then discuss the factors under each theme.

### 4.1 Collaborative Culture

A collaborative culture is unequivocally a core theme regarding the success of the DevOps environment [17]. Three factors that make up this Collaborative Culture are Collaboration, Communication and Sharing.

**Success Factor 1: Collaboration.** Collaboration is critical in a DevOps environment because it is the essence of DevOps itself as it focuses on working together, and the exchange of knowledge and experience between teams. DevOps is all about connecting Development and Operations, which entails collaboration [23]. The following 13 authors identified collaboration in this SLR as one of the essential aspects of DevOps environments [2], [6], [9], [10], [12], [14], [17], [20], [23], [28], [29], [32], [33]. This is because numerous benefits have been seen when a DevOps environment has a strong culture of collaboration. These advantages include enhanced development and operations processes, as well as improved quality of software services and products [9].

**Success Factor 2: Communication.** Communication between the operations and development teams is essential for DevOps to succeed [2]. A DevOps environment is essentially made up of a cross-functional team. Thus, for the teams to work together effectively, information within and between the teams needs to be shared effectively and frequently. The presence of quality communication helps establish a collaborative culture which contributes to a successful DevOps environment. [2] emphasise that communication is crucial for DevOps to succeed because not sharing important information may have a significant negative impact, therefore, managers must impose regular communication. [21] cited communication as a critical success factor for DevOps improvement. [21] further posits that effective communication is the most important success factor during DevOps implementation in a software organisation.

**Success Factor 3: Sharing.** Sharing is prevalent in a thriving DevOps environment because it has a tremendous positive impact on its process and output. Sharing in DevOps means the joint use of tools, practices, processes, and knowledge [33]. [1], [4-9], [11], [17], [22], [28] states how sharing tools, practices, and processes impact the DevOps environment. [29] identified sharing resources as one of the success factors of collaboration in a DevOps context. [33] point out how sharing failures and successes in DevOps practices helps team members improve quicker by learning from others.

### 4.2 Organisational Aspects

Certain aspects within a software organisation influence the success of the DevOps environment. These aspects include Team Role Setting, Culture, Training and Customer Involvement.

**Success Factor 4: Team Role Setting.** Team Role Setting is one of the factors that are still blurry in many DevOps environments, but when team roles are set, it impacts the success of these environments. This factor has to do with the precise definition of roles and responsibilities within a DevOps team. This definition of roles is crucial because the observed trend in DevOps is the reorientation of roles and responsibilities between software development and operations teams [19]. Clear team roles have several benefits. Some of the benefits are; positive effects on team members' performance [20], helping practitioners in recognising their crucial role in DevOps processes [2] and assisting in aligning the motivations of the key roles involved in software delivery [20]. [30] also identified clear role setting as one of the success factors of collaboration in the DevOps context.

**Success Factor 5: Organisational Culture.** Organisational Culture is an integral part of the DevOps environment, and changing the organisational culture to have the attributes of DevOps culture is a decisive success factor. DevOps features culture as one of the vital aspects to safeguard the success of DevOps adoption [28]. Even though DevOps have varying definitions, most authors agree that culture is the epitome of DevOps. The popular interpretation is that DevOps is about culture [24], and culture change is seen as the beginning of DevOps [23]. In an organisation, you will find team culture and organisational culture. Team culture flows from the organizational culture [24]. Therefore, it is integral that the organisational culture represents a DevOps culture.

**Success Factor 6: Training.** Training is one of the more overlooked success factors for multiple reasons. However, although it was cited the least in this SLR (tied with continuous delivery), its impact is undoubtedly noteworthy. Appropriate training is required to grasp the DevOps and DevOps environment concepts well [33]. DevOps requires supplementary technical skills from both software developers and operators since operators must pick up some skills from developers and vice versa [10]. Therefore, training is imperative for DevOps success and must be considered a priority for organisations.

**Success Factor 7: Customer Involvement.** Customer Involvement was identified as a success factor for software organisations' DevOps environment due to the customer feedback's impact on software product improvement. The term customer involvement is extensive and has several definitions. [31] describes customer involvement as "*preparing and receiving customer input, establishing a customer sample group, and delivering feature growth*." Through verified performance improvement, [31] identified customer involvement as a critical success factor during the adoption of continuous deployment and continuous delivery in the context of DevOps. In a study by [21], 87% of the respondents considered 'Customer feedback to improve development' the second most crucial success factor in DevOps adoption.

### 4.3 Tooling and Technology

The selection of tools and use of technology in DevOps significantly affects the success of a DevOps environment [6]. This section discusses how three factors, Automation, Measurement, and Tooling, contribute to a DevOps environment's success.

**Success Factor 8: Automation.** Automation is referred to as one of the key aspects of DevOps because of the advantages it brings to a DevOps environment. Making repetitive manual tasks automated has contributed to success in DevOps environments within a software organisation. [23] identified six concepts of automation in DevOps. 1) Build automation, 2) recovery automation, 3) deployment automation, 4) monitoring automation, 5) test automation and 6) infrastructure automation (also referred to as infrastructure as code). However, three concepts of automation that are widely agreed upon in our SLR as having benefits that contribute to success in a DevOps environment are test automation, deployment automation, and infrastructure automation.

*Deployment automation* automates software deployments and rollbacks of software build to be transferred to development, testing, acceptance, and production environments quicker [7]. Deployment automation reduces user errors and increases deployment speed [10].

*Test automation* is executing tests and repeating activities using software [33]. Test automation reduces manual testing and accelerates the testing process significantly [10]. In an experiment conducted by [26], they observed that test automation strongly affected deployment speed.

*Infrastructure automation* provides infrastructure elements as code and configuration of application servers using pre-set automation logic [10]. Automation improves the consistency of infrastructure, team productivity, and repeatability of activities [9].

In addition, deployment, test, and infrastructure automation form part of the DevOps continuous practices, discussed in section 4.4.

**Success Factor 9: Tooling.** Tooling in DevOps is posited to affect the success of a DevOps environment because it is not just about using tools but using the right tools. Selecting the wrong tools might negatively influence the productivity of work [3]. DevOps tools need to support DevOps practices as this is the primary purpose of the tools and how their contribution to success is realised. Tools are used to build DevOps capabilities and collaboration [14]. DevOps tools assist in monitoring, implementing, deploying, automating, and analysing each process in the DevOps journey [3].

**Success Factor 10: Measurement.** Measurement in DevOps environments is important because it influences the software development process. Measurement focuses on specifying production metrics to monitor and evaluate the performance of a process in software development and operations [19]. As indicated, this factor's significance stems from the impact of having metrics and a baseline to assess performance in a software organisation's DevOps environment. The sooner an organisation applies measurement, the sooner they establish a baseline to evaluate improvement [8].

### 4.4 Continuous Practices

The practices done constantly in a DevOps environment are considered vital to the environment's success; these are Continuous Practices. Continuous Practices are Continuous Integration, Continuous Deployment, Continuous Monitoring, and Continuous Experimentation.

**Success Factor 11: Continuous Integration.** Continuous Integration is a straightforward success factor in a DevOps environment, as DevOps demands a need for this [27]. It is defined by [31] as the part that deals with code compilation, acceptance testing, unit testing, confirming code coverage, examining compliance to coding standards, and deployment package building. Continuous integration is an incremental process that prevents last-minute integration issues and is about achieving one of the most essential values of DevOps, which is speed [14]. In addition, continuous integration reduces the risk and cost associated with software releases [33].

**Success Factor 12: Continuous Delivery.** Continuous Delivery is an integral part of the DevOps environment, as it allows DevOps teams to automate their software delivery process up to the point just before it is deployed [34]. Continuous delivery as a DevOps practice correlates to performance [14], and it has been argued that measurement contributes to continuous delivery [17], [30]. An essential benefit of this practice, as mentioned by [14], [17], and [30], is how it creates an environment for quality outcomes. In other words, by implementing the continuous delivery practice, quality can be seen as a by-product.

**Success Factor 13: Continuous Deployment.** Continuous Deployment allows DevOps teams to automatically deploy new software features onto a production environment [19], [35]. No manual human intervention is needed to enable the software to be deployed. [19] states that the popularity of DevOps results from the emergence of continuous deployment, and [30] suggests this practice reduces the costs of detecting defects due to faster feedback from users. Therefore, it is no surprise that seven of the included studies mentioned it as a factor. Continuous development is posited to contribute to DevOps success within software organisations.

**Success Factor 14: Continuous Monitoring.** Continuous Monitoring is a factor whose contribution to the DevOps environment hinges on the presence of another factor, namely, measurement. However, monitoring is a continuous and ongoing activity, and therefore it is classified as a continuous practice. Once metrics in measurement have been set, continuous monitoring commences. Continuous improvement in DevOps is enabled by constantly monitoring processes and measuring them against metrics [7].

**Success Factor 15: Continuous Experimentation.** Continuous Experimentation is an infrequently used term but was identified in eight of the selected literature as having a positive impact on a DevOps environment that could contribute to success. Continuous experimentation is defined by [22] as the continual testing of propositions to determine the benefits for the customer and the software organisation. An in-depth analysis by [19] of their primary studies revealed continuous and rapid experimentation as one of the recurring factors in continuous deployment. [24] identified experimentation as one of the characteristics of a healthy DevOps environment. These findings are of great importance because continuous experimentation is one of the essential practices in a DevOps environment.

The systematic literature review identified factors that are posited to contribute to the success of DevOps environments. Collaboration and Organisational Culture were the most cited factors, which can infer that they are the most important. However, this research did not prioritise the factors because it is beyond the scope of this study.

It was observed that the influence of some factors was at distinct phases of software development, while others influenced the entire software development process. The factors associated with Organization Aspects and Collaborative Culture influence the DevOps environment at every phase of software development, from initial team setup to product delivery. In contrast, factors related to Tooling and Technology and Continues Practices have more impact once software development begins.

## 5      Conclusion

A systematic literature review (SLR) was conducted to identify the factors that contribute to the success of a software organisation's DevOps environment. Thirty-three articles were selected from five data sources for data extraction and synthesis. The data synthesis was conducted using a thematic analysis following the data extraction. For the first iteration of data synthesis, 16 factors were identified. After the second iteration of data synthesis, the factors were refined to 13. After the final iteration of data synthesis with the help of the NVivo tool, two additional factors were included, settling on 15 factors posited to contribute to the success of a DevOps environment within software organizations. As a result of this study, an initial conceptual model was developed. The model developed provides the mapping of the identified factors to constructed themes.  The factors were classified into four themes to answer the research question - Collaborative Culture (3 factors), Organisational Aspects (4 factors), Tooling and Technology (3 factors), and Continues Practices (5 factors). The research aims were met, aiding in answering the research question.

This study's limitations are that English papers were only considered for this SLR; thus, any information that might have been relevant to the research question but was written in another language is excluded. Secondly, only five data sources were used to search for the literature for this SLR; therefore, relevant literature from other databases might have been missed. Thirdly, some factors were excluded or merged into others due to a lack of a clear definition or information indicating how they contribute to success. The implication might be that some standalone factors in a DevOps environment were merged with others in this study. Fourth, the search terms used in the search string may not be of adequate rigour, which could have resulted in missed factors and themes.

Researchers and practitioners could use the model derived from this SLR to discover the critical success factors in a DevOps environment. Furthermore, another study could investigate the theme's and factor's level of contribution towards DevOps Environment success. In addition, a similar study could be undertaken to confirm (or debunk) the success factors and themes in this study.

15